\documentclass[12pt]{article}
\usepackage{cite} 
\usepackage{amsfonts,latexsym}
\usepackage{epsfig}
\begin{document}
\newcommand{\de}{\delta}\newcommand{\ga}{\gamma}
\newcommand{\e}{\epsilon} \newcommand{\ot}{\otimes}
\newcommand{\be}{\begin{equation}} \newcommand{\ee}{\end{equation}}
\newcommand{\ba}{\begin{array}} \newcommand{\ea}{\end{array}}
\newcommand{\beq}{\begin{equation}}\newcommand{\eeq}{\end{equation}}
\newcommand{\tmod}{{\cal T}}\newcommand{\amod}{{\cal A}}
\newcommand{\bemod}{{\cal B}}\newcommand{\cmod}{{\cal C}}
\newcommand{\dmod}{{\cal D}}\newcommand{\hmod}{{\cal H}}
\newcommand{\s}{\scriptstyle}\newcommand{\tr}{{\rm tr}}
\newcommand{\einsop}{{\bf 1}}
\def\R{\overline{R}} \def\doa{\downarrow}
\def\dag{\dagger}
\def\ve{\epsilon}
\def\si{\sigma}
\def\ga{\gamma}
\def\nn{\nonumber}
\def\le{\langle}
\def\re{\rangle}
\def\lt{\left}
\def\rt{\right}
\def\dwn{\downarrow}   
\def\up{\uparrow}
\def\j{\mathcal{J}}
\def\dag{\dagger}
\def\bea{\begin{eqnarray}}
\def\eea{\end{eqnarray}}
\def\p{\tilde{p}}
\def\q{\tilde{q}}
\def\H{\overline{H}}
\newcommand{\reff}[1]{eq.~(\ref{#1})}

\title{The two-site Bose--Hubbard model}

\author{ J. Links$^1$, A. Foerster$^2$,  A. Tonel$^2$ and  G. Santos$^2$
\vspace{1.0cm}\\
$^{1}$ Centre for Mathematical Physics, School of Physical Sciences \\ 
The University of Queensland, Queensland, 4072, Australia
\vspace{0.5cm}\\
$^{2}$ Instituto de F\'{\i}sica da UFRGS \\ 
Av. Bento Gon\c{c}alves 9500, Porto Alegre, RS - Brazil}

\maketitle

\begin{abstract}
The two-site Bose--Hubbard model is a simple model used to study
Josephson tunneling between two Bose--Einstein condensates. 
In this work we give an overview of some mathematical aspects of
this model. Using a classical analysis, 
we study the equations of motion and the level curves
of the Hamiltonian. Then, the quantum dynamics 
of the model is investigated   
using direct diagonalisation of the Hamiltonian. 
In both of these analyses, 
the existence of a threshold coupling between a delocalised 
and a self-trapped phase 
is evident, in qualitative agreement with experiments.
We end with a discussion of the exact solvability of the model via the
algebraic Bethe ansatz. 

\end{abstract}

\vspace{1cm}

{\it This work is dedicated to the memory of Daniel Arnaudon} 

\vfil\eject

\section{Introduction}
The phenomenon of Bose--Einstein condensation, while predicted
long ago \cite{bose,eins}, is nowadays responsible for many 
current perspectives on the potential applications of quantum
behaviour in mesoscopic systems. 
This point of view has arisen with the experimental
observation of condensation in systems of ultracold dilute alkali
gases, realised by several research groups using magnetic traps
with various kinds of confining geometries \cite{cw,ak}.
These types of experimental
apparatus open up the possibility for studying  
quantum effects, such as Josephson tunneling and self-trapping
\cite{wwcch,albiez}, in a macroscopic setting.

{}From the theoretical point of view, the two-site Bose--Hubbard model 
(see eq. (\ref{ham}) below), also known as the 
{\it canonical Josephson Hamiltonian} \cite{l01}, has been a useful model in 
understanding tunneling phenomena. The simplicity of the model means
that it is amenable to detailed mathematical analysis, as we will discuss below.
However despite this apparent simplicity, the
Hamiltonian captures the essence of competing linear and non-linear
interactions, leading to interesting, non-trivial behaviour. 

The Hamiltonian is given by 
\bea
H&=& \frac {k}{8}  (N_1- N_2)^2 - \frac {\Delta \mu}{2} (N_1 -N_2)
 -\frac {\cal {E} _J}{2} (b_1^\dagger b_2 + b_2^\dagger b_1).
 \label {ham} \eea
 where $b_1^\dagger, b_2^\dagger$ denote the single-particle creation
 operators in the two wells and  $N_1 = b_1^\dagger b_1,
 N_2 = b_2^\dagger b_2$ are the corresponding
 boson number operators. The total boson number $N_1+N_2$
 is conserved and set to the fixed value of $N$.
 The coupling $k$ provides the strength of the scattering interaction between 
 bosons,
 $\Delta \mu$ is the external potential and ${\cal E}_J$ is the coupling
 for the
 tunneling.
 The change ${\cal E}_J\rightarrow -{\cal E}_{J}$ corresponds to the
 unitary
 transformation $b_1 \rightarrow b_1,\,b_2\rightarrow -b_2$, while
 $\Delta \mu \rightarrow -\Delta \mu$ corresponds to $b_1
 \leftrightarrow b_2$.
 Therefore we will restrict our analysis to the case of
 ${\cal E}_J,\,\Delta\mu\geq 0$.
 For $k>0$, following \cite{l01} it is useful to divide the parameter
 space into three regimes; viz. Rabi ($k/{\cal E}_J<< N^{-1}$),
 Josephson ($N^{-1}<<k/{\cal E}_J<<N$) and Fock ($N<<k/{\cal E}_J $).
 For these three regimes, there is a correspondence between
 (\ref{ham}) and the motion of a pendulum \cite{l01}. In the Rabi and
 Josephson
 regimes this motion is semiclassical, in contrast to the Fock regime.
 For both the Fock and Josephson regimes the analogy corresponds to a
 pendulum with fixed length, while in the Rabi regime the length varies.

In the present work, we give an overview of some of the mathematical
aspects of (\ref{ham}). We 
undertake an analysis of the classical and the quantum
dynamics of the system, and discuss how  
the system exhibits a threshold coupling, originally identified in \cite{milb},
about which the dynamics abruptly changes. 
Below this threshold point the dynamics is delocalised, 
while above it the dynamics turns out to be localised 
(macroscopic self-trapping). This can be seen at both the classical and
quantum level. 
The result is in qualitative agreement with experimental results \cite{albiez}.
We conclude by giving an outline of the algebraic Bethe ansatz solution 
of (\ref{ham}).
  
\section{Classical dynamics}

First we study a classical analogue  
of  the model. 
Let $N_j,\,\theta_j,\,j=1,\,2$ be
quantum variables satisfying the canonical relations
$$[\theta_1,\,\theta_2]=[N_1,\,N_2]=0,~~~~~[N_j,\,\theta_k]=i\delta_{jk}I.$$
Using the fact that
$$\exp(i\theta_j)N_j=(N_j+1)\exp(i\theta_j) $$
we make a change of variables from the operators
$b_j,\,b_j^\dagger,\,j=1,\,2$ via
$$b_j=\exp(i\theta_j)\sqrt{N_j},
~~~b_j^\dagger=\sqrt{N_j}\exp(-i\theta_j) $$
such that the Heisenberg canonical commutation relations are preserved.
Now define the variables
$$z=(N_1-N_2)/N $$
$$\phi=N(\phi_1-\phi_2)/2 $$
where $z$ represents the fractional occupation difference 
(or the {\it imbalance}) and 
$\phi$ the phase difference. In the classical limit where 
$N$ is large, but still finite,  
we may equivalently consider the Hamiltonian 
\cite{rsfs,ks}
\begin{equation}
H(z,\phi)= \frac{{\cal E}_JN}{2} \left(\frac{\lambda}{2}z^2-\beta z 
- \sqrt{1-z^2}\cos({2\phi}/{N})\right)
\label{hsc}
\end{equation}
where $$\lambda=\frac{kN}{2{\cal E}_J},~~~~~
\beta=\frac{\Delta\mu}{{\cal E}_J} $$
and $(z,\,\phi)$ are canonically conjugate variables.
We note the Hamiltonian (\ref{hsc}) obeys the symmetries 
 \begin{eqnarray}\left.H\left(z,\phi\right)\right|_{\lambda,\beta}
&=&-\left.H\left(z,\phi+N\pi/2\right)\right|_{-\lambda,-\beta} \nonumber \\
 \left.H\left(z,\phi\right)\right|_{\lambda,\beta}
&=&\left.H\left(-z,\phi\right)\right|_{\lambda,-\beta}. \label{symm}
\end{eqnarray}

The classical dynamics is given by Hamilton's equations of motion
\begin{eqnarray}
\dot{\phi}&=& \frac{\partial H}{\partial z}=\frac{{\cal E}_JN}{2} \left(
\lambda z - \beta +\frac{z}{\sqrt{1-z^2}}\cos({2\phi}/{N})\right)\nonumber \\
\dot{z}&=&-\frac{\partial H}{\partial \phi}=-{\cal E}_J 
\left(\sqrt{1-z^2}\sin({2\phi}/{N})\right).
\label{eqmov}
\end{eqnarray}
Now we study the fixed points of the Hamiltonian
  (\ref{hsc}),
  determined by the condition
  $\dot{z}=\dot{\phi}=0$.
This leads to the following classification:

  \begin{itemize}
  \item $\phi=0$ and $z$ is a solution of

  \begin{equation}
  \lambda z -\beta = - \frac{z}{\sqrt{1-z^2}}
  \label{teq}
  \end{equation}
  which has a unique real solution for $\lambda> 0$.

  \item $\phi={N\pi}/{2}$  and $z$ is a solution of
  \begin{equation}
  \lambda z - \beta= \frac{z}{\sqrt{1-z^2}}.
  \label{teq1}
  \end{equation}
  This equation has either one, two or three real solutions for $\lambda> 0$. \\
  \end{itemize}

\begin{figure}[ht]
\begin{center}
\epsfig{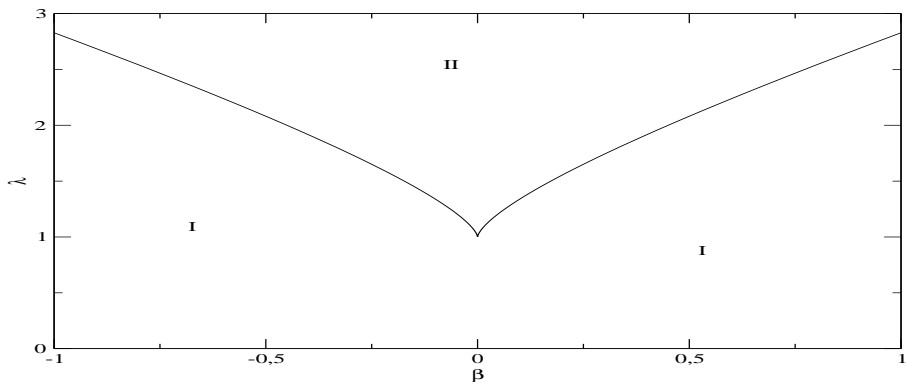}
\caption{Coupling parameter space diagram identifying the different
types of solutions for
equation (\ref{teq1}). In region I there is just one  solution for 
$z$,  a local maximum. 
In region II there are  three solutions for
$z$, two local maxima and a saddle point.
The boundary separating regions I and II is given by (\ref{boundary}). 
}
\label{fig3}
\end{center}
\end{figure}

From eq. (\ref{teq1}) we can determine that there are fixed point
bifurcations for certain choices of the coupling parameters.
These bifurcations allow us to divide the coupling parameter space in
two regions.  
Setting $f(z)=\lambda z-\beta$ and 
$g(z)=z({1-z^2})^{-1/2}$, the boundary between the regions occurs when
$f(z)$ is the tangent line to $g(z)$ at some value $z_0$. A standard
analysis shows this occurs when $\lambda=g'(z_0)=(1-z_0^2)^{-3/2}$.
Requiring $f(z_0)=g(z_0)$ then yields the following relationship   
\begin{equation}
\lambda=(1+\left|\beta\right|^{2/3})^{3/2}
\label{boundary}
\end{equation}
determining the boundary. This is depicted in Fig. \ref{fig3}.


\vspace{1.0cm}

This leads us to the following classification: 
\begin{itemize}
\item $ 0 < \lambda<1$:
For any value of $\beta$ there is just one real solution, for
which the Hamiltonian attains a local maximum. \\

\item  $\lambda >1$: Here transition couplings $\pm\beta_{0}$ appear,
which can be seen from Fig. \ref{fig3}.
For $\beta \in (-\beta_0 ,\beta_{0})$, the equation has two locally
maximal fixed points and one saddle point, while for $\beta>\beta_{0}$ or  $\beta <
-\beta_{0}$
 the equation has just one real solution, a locally
  maximal fixed point. 
     \end{itemize}
      
We remark that in the absence of the external potential
       $(\Delta\mu=\beta=0)$ the transition value is given by
	 $\lambda_0=1$. Using the symmetry relation (\ref{symm}) we can deduce that for 
the attractive case $\lambda<0$, $\lambda_0=-1$ is the coupling
marking a bifurcation between a locally minimal fixed point (for $\lambda>-1$) and
two locally minimal fixed points and a saddle point (for $\lambda<-1$). This is a supercritical
pitchfork bifurcation of the classical ground state. The results of \cite{hines} predict that the ground-state 
entanglement, as measured by the von Neumann entropy, is maximal at this coupling. The numerical results of \cite{pd} confirm this.

\begin{figure}[ht]
\begin{center}
\epsfig{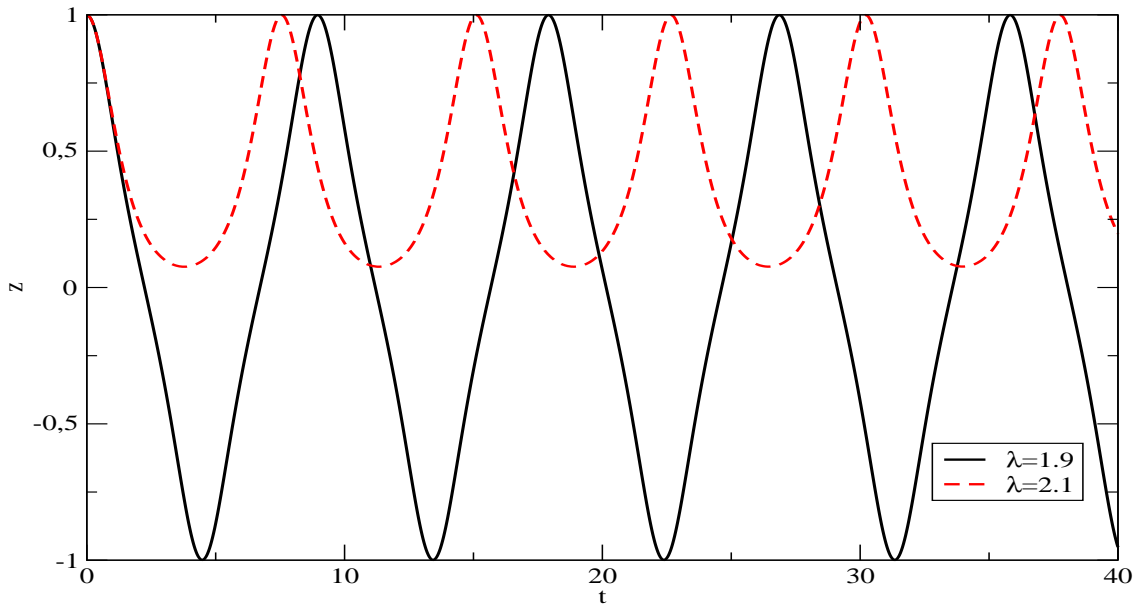}
\caption{Time evolution for the imbalance $z$. The solid line is for 
$\lambda=1.9$, 
while the dashed curve is for $\lambda=2.1$. Here we are using $N=100$, ${\cal E}_J=1$ 
and the initial conditions $z(0)=1$, $\phi(0)=0$. The threshold coupling
occurs at $\lambda_c=2$.}
\label{cd}
\end{center}
\end{figure}

Next we look at the dynamical evolution. 
In that which follows we will consider the 
equations (\ref{eqmov}) in the absence of the external field 
($\Delta \mu=0$ or, equivalently, $\beta=0$). 
An analysis including the effect of this term can be found in references 
\cite{ours,buon}.
We integrate (\ref{eqmov}) to find the time evolution for the 
imbalance $z$, using the initial condition $z(0)=1,\,\phi(0)=0$. 
By plotting $z$ against the time, it is evident that
there is a threshold coupling $\lambda_c=2$ separating two different 
behaviours in the 
classical dynamics, as can be seen in Fig. \ref{cd}: 
\begin{itemize}
\item[(i)] For $\lambda<2$ the  system oscillates 
between $z=-1$ and $z=1$. Here the evolution is delocalised;  
\item[(ii)] For $\lambda>2$ the system oscillates between $z=0$ and $z=1$. 
Here the evolution is localised. 
\end{itemize}
The threshold occuring at $\lambda_c=2$ 
was first observed in \cite{milb}. 

\vspace{1.0cm}

To help visualise the classical dynamics, 
it is useful to plot the level curves (constant energy curves)  
of the Hamiltonian (\ref{hsc}) in phase space. 
Given an initial condition $(z(0), \phi(0))$, 
the system follows a trajectory along the level curve $H(z(0),\phi(0))$.
In Fig. {\ref{level2} we plot the level curves for different values of 
$\lambda$ ($\lambda=1.5$ on the left and $\lambda=2.5$ on the right), 
where we take $2\phi/N\in[-\pi,\,\pi]$.
We can observe clearly two distinct scenarios:
\begin{itemize}
\item $ \lambda > 2$: Here we see that for the orbit with initial condition $z_0=1,\,\phi(0)=0$, $\phi$ increases monotonically (running phase mode). The evolution of $z$ is bounded in the interval $[0,1]$, leading to localisation
(self-trapping). 
\item $ \lambda < 2$: Here we see that for the orbit with initial condition $z(0)=1,\,\phi(0)=0$, the evolution of 
$\phi$ is oscillatory and bounded in the interval $(-N\pi/2,\,N\pi/2)$. The evolution of $z$ is not bounded, leading to delocalisation. 
\end{itemize}

\begin{figure}[ht]
\begin{center}
\begin{tabular}{cc}
            &             \\
    (a)& (b)    \\
\epsfig{file=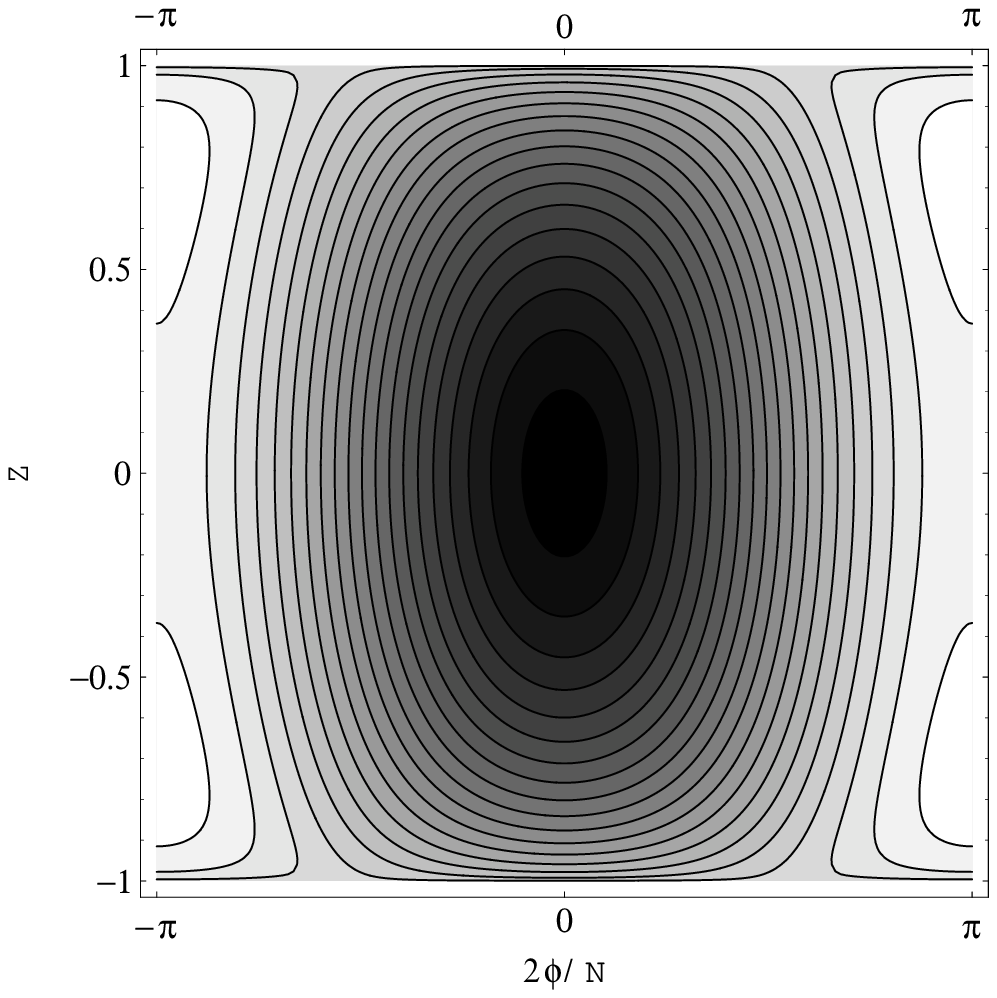,width=6cm,height=6cm,angle=0}&            
\epsfig{file=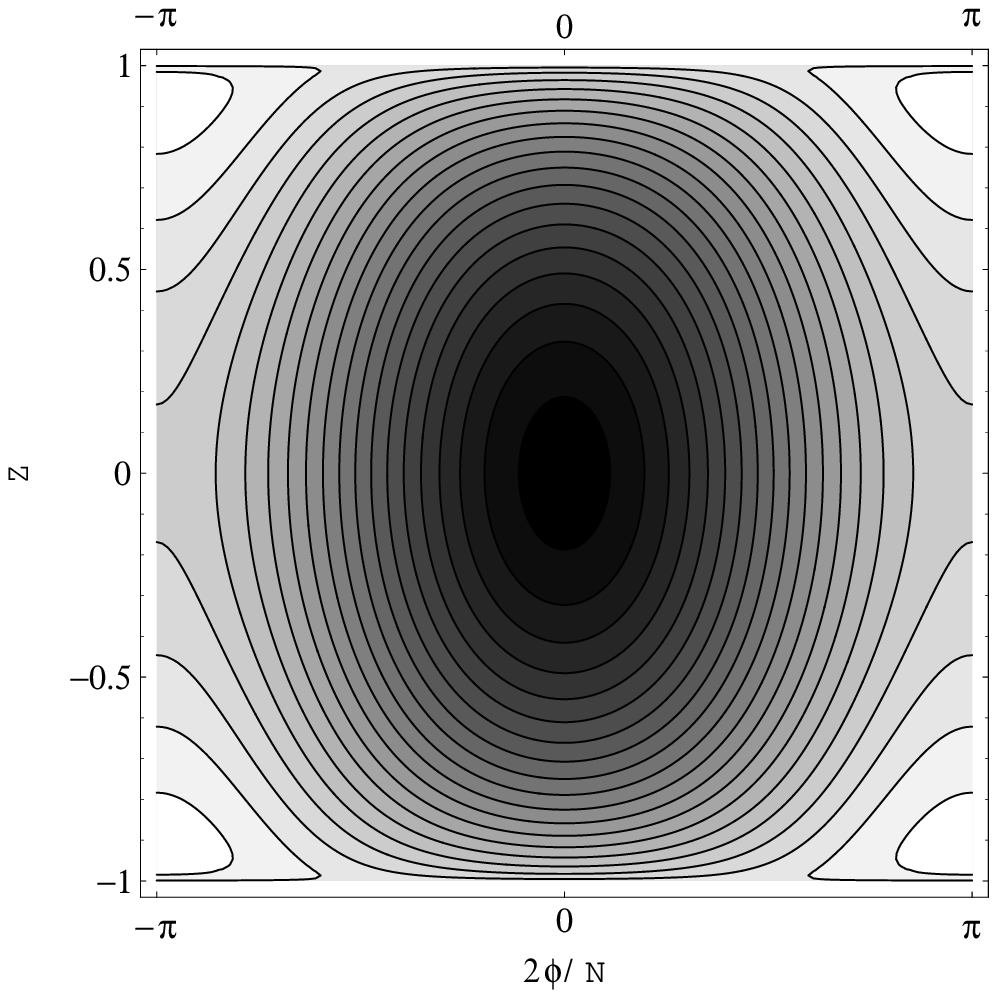,width=6cm,height=6cm,angle=0} \\
\end{tabular}
\end{center}
\caption{Level curves of the Hamiltonian (\ref{hsc}) (a) for $\lambda=1.5$ (below the 
threshold point) and (b) for $\lambda=2.5 $ (above the threshold point). 
We are using $N=100$ and ${\cal E}_J=1 $.
Above the threshold coupling running phase modes occur leading to localised evolution of $z$. Below the threshold
coupling the evolution of $z$ is delocalised.} 
\label{level2}
\end{figure}

The threshold coupling $\lambda_c=2$ (or  
$k/{\cal E}_\j=4/N$, in terms of the original variables)
separates two distinct dynamical behaviours.
This value for the threshold 
between delocalisation and self-trapping also occurs for the quantum 
dynamics, as we will show in the next section.


\section{Quantum dynamics}

We will investigate the quantum dynamics of the 
Hamiltonian in the absence of the external potential
$(\Delta \mu=0$)
using the exact diagonalisation method. 
It is well known that the  
time evolution of any state is determined 
by $|\Psi(t) \rangle = U(t)|\phi_0 \rangle$, 
where $U$ is the temporal evolution operator given by 
$U(t)=\sum_{m=0}^{M}|m \rangle \langle m|\exp(-i E_m t)$,  
$|m\rangle$ is an eigenstate with energy $E_m$ and $|\phi_0 \rangle$ 
represents the initial state.
Using these expressions we can compute the expectation value of the 
relative number of particles
\begin{equation}
\langle(N_1-N_2)(t)\rangle=\langle \Psi (t)|N_1-N_2|\Psi (t)\rangle. 
\end{equation} 
From Fig. \ref{fig.2} it is seen that the qualitative behaviour in
each region does not depend on the
number of particles. 
We find that in the interval
$k/{\cal E}_\j\in [1/N^2,1/N]$ (close to the Rabi regime) 
the collapse and revival time takes the constant value
$t_{cr}=4\pi$\footnote{The ratio $k/{\cal E}_\j=1/N^2$ means that
we are using $k=1$ and ${\cal E}_\j=N^2$ and similarly for the
other cases. }
In the interval between $k/{\cal E}_\j=1/N$ and $k/{\cal
E}_\j=1$ the system undergoes a transition from oscillations which vary between positive and negative values
of $\left<N_1-N_2\right>$ 
(delocalised) to one where $\left<N_1-N_2\right>$ is close to $N$ (self-trapping). 

\vspace{1.00cm} 
\begin{figure}[ht]
\begin{center}
\epsfig{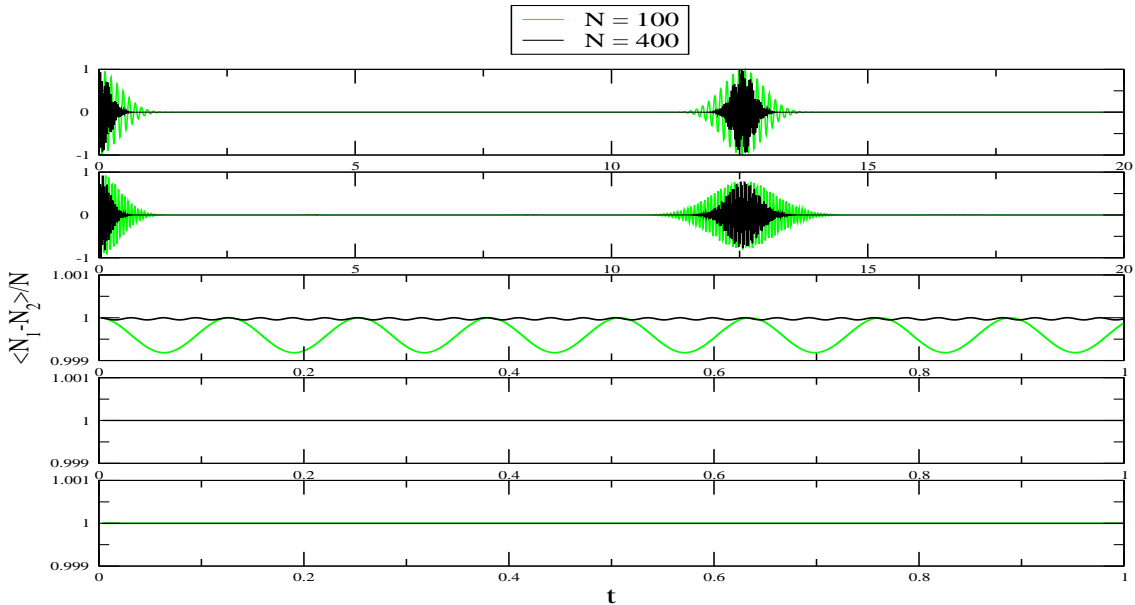}
\caption{Time evolution of the expectation value for the relative
number of particles
for different ratios of the coupling
$k/{\cal E}_\j$ from the top (Rabi regime) to the bottom (Fock
regime):
$k/{\cal E}_\j= 1/N^2,1/N,1,N,N^2$ for $N=100, 400$ and the initial
state is $|N,0\rangle $.}
\label{fig.2}
\end{center}
\end{figure}

Now we focus in more detail the time evolution of the expectation value of the
relative number of particles in the interval $k/{\cal E}_\j\in [1/N,1]$
In Fig. \ref{fig.3} we present the case $N=100$: we observe 
the evolution of the dynamics from a collapse and revival sequence 
for $k/{\cal E}_\j<4/N$, through  the self-trapping
transition at $k/{\cal E}_\j=4/N$, and toward small amplitude
harmonic oscillations in the imbalance of the localised state when
$k/{\cal E}_\j=1$. It is also interesting to observe in the localised
phase $k/{\cal E}_\j>4/N$ 
the re-emergence of a collapse and revival sequence.
Further increases in $k/{\cal E_\j}$
lead to a decaying of the collapse and revival sequence toward
harmonic oscillations which occur at $k/{\cal E}_\j=1$. 
A more detailed investigation, using another initial conditions can be found in ref. \cite{our}.

\vspace{1.0cm}
\begin{figure}[ht]
\begin{center}
\epsfig{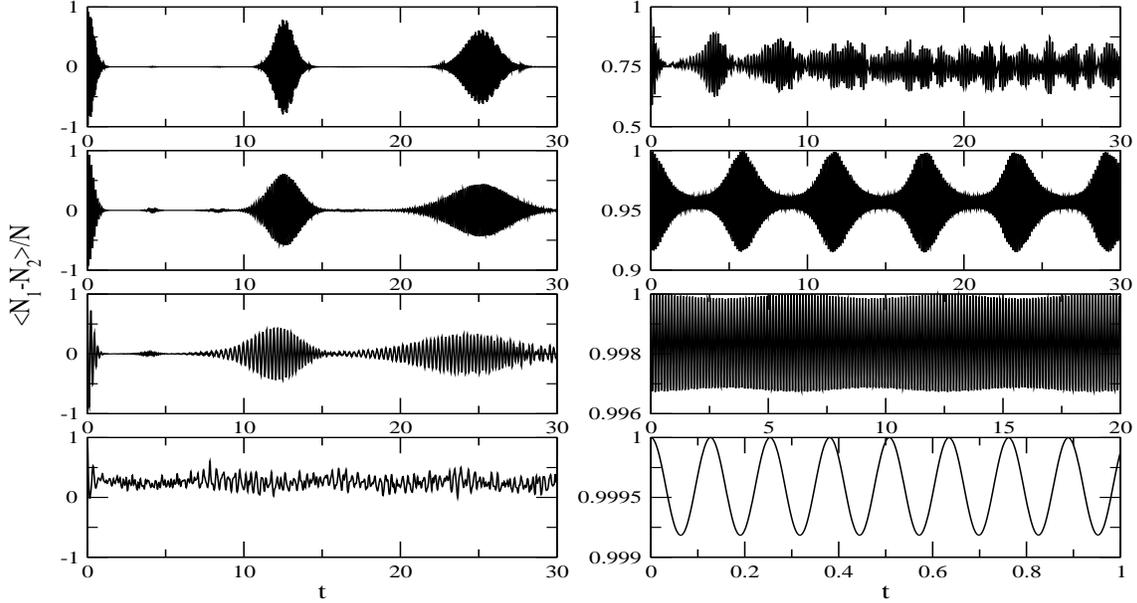}
\caption{Time evolution of the expectation value between
$k/{\cal E}_\j=1/N$
and  $k/{\cal E}_\j= 1$. On the left, from the top to the bottom
$k/{\cal E}_\j= 1/N,2/N,3/N,4/N$ and on
the right, from the top to the bottom  $k/{\cal E}_\j=5/N,10/N,50/N,1$, where
 $N=100$ and  the initial state is $|N,0\rangle $.}
\label{fig.3}
\end{center}
\end{figure}

{}From the above picture it is clear that the threshold coupling
$k/{\cal E}_\j=4/N $ predicted by the classical analysis, representing 
the boundary between a delocalised evolution ($k/{\cal E}_\j < 4/N $) 
and self-trapped evolution ($k/{\cal E}_\j> 4/N $), 
also holds for the quantum dynamics.

\section{Exact Bethe ansatz solution}

In this final section we briefly discuss the exact
Bethe ansatz solution of (\ref{ham}). More details can be found in \cite{jrlreview,angela}.
We begin with the $SU(2)$-invariant $R$-matrix, depending on the
spectral parameter $u$:
\begin{equation}
R(u) =  \left ( \begin {array} {cccc}
1&0&0&0\\
0&b(u)&c(u)&0\\
0&c(u)&b(u)&0\\
0&0&0&1\\
\end {array} \right ),
\label{r}
\end{equation}
with $b(u)=u/(u+\eta)$ and
$c(u)=\eta/(u+\eta)$.
Above, $\eta$ is an arbitrary parameter, to be chosen later.
It is easy to check that $R(u)$ satisfies the Yang--Baxter equation
\begin{equation}
R _{12} (u-v)  R _{13} (u)  R _{23} (v) =
R _{23} (v)  R _{13}(u)  R _{12} (u-v).
\label{ybe}
\end{equation}
Here $R_{jk}(u)$ denotes the matrix  acting non-trivially on the
$j$-th and $k$-th spaces and as the identity on the remaining space.

Next we define the Yang--Baxter algebra $T(u)$,
\begin{equation}
T(u)=\left(\matrix{A(u)&B(u)\cr
C(u)&D(u)\cr}\right)
\label{mono}
\end{equation}
subject to the constraint
\begin{equation}
R_{ab}(u-v) T_a(u) T_b(v)=
T_b(v) T_a(u)R_{ab}(u-v).
\label{yba}
\end{equation}
We may choose the following realization for the Yang--Baxter algebra
\begin{equation}
\pi (T_a(u)) = L_{a1}(u + \omega )L_{a2}(u - \omega),
\end{equation}
written in terms of 
the Lax operators \cite{jrlreview}
\begin{equation}
L_i(u)=\left(\matrix{u+\eta N_i&b_i\cr
b_i^{\dagger}&\eta^{-1}\cr}\right)\;\;\;\;\;\; i=1,2.
\end{equation}
Since $L(u)$ satisfies the relation
\begin{equation}
R_{ab}(u-v)L_{ai}(u)L_{bi}(v)=L_{bi}(v)L_{ai}(u)R_{12}(u-v), ~~~i=1,\,2   
\end{equation}
it is easy to check that the Yang-Baxter algebra (\ref{yba})
is also obeyed.

Finally, defining the transfer matrix through
\begin{equation}
t(u) = \pi ( {\rm tr}_a T_a(u)) = \pi ( A(u)+D(u) )
\label{tm}
\end{equation}
it follows from (\ref{yba}) that the
transfer matrix commutes for different values of the spectral
parameters;
i.e., the model is integrable.
Now it is straightforward to check that the
Hamiltonian (\ref{ham})
is related with the
transfer matrix  $t$ (\ref{tm}) through
$$
H=-\kappa \left (t(u) -\frac{1}{4} (t'(0))^2-
u t'(0)-\eta^{-2}
+\omega^2 -u^2\right),
$$
where the following identification has been made for the coupling
constants
\begin{equation}
\frac {k}{4} =  \frac {\kappa \eta^2}{2}, ~~~
\frac {\Delta \mu}{2} =  -\kappa \eta \omega, ~~~
\frac {\cal {E}_J}{2} =  \kappa . \nonumber
\end{equation}

We can apply the algebraic Bethe ansatz method, using the  Fock
vacuum as the pseudovacuum, to
find the Bethe ansatz equations (BAE)
\beq
\eta^2 (v^2_i -\omega^2)=
\prod ^N_{j \neq i}\frac {v_i -v_j - \eta}{v_i -v_j +\eta}
\label{becbae} \eeq
and the energies of the Hamiltonian (see for example
\cite{jrlreview,angela})
\begin{eqnarray}
E&=&-\kappa\left(\eta^{-2}\prod_{i=1}^N(1+\frac{\eta}{v_i-u})
-\frac{\eta^2N^2}{4} -u\eta N-u^2
\right. \nonumber \\
&&~~~~~~\left.
-\eta^{-2}+\omega^2+(u^2-\omega^2)\prod_{i=1}^N(1-\frac{\eta}{v_i-u})
\right).     \label{becnrg}
\end{eqnarray}
Surprisingly this expression is independent of the spectral parameter $u$, which
can be chosen arbitrarily.

Using the Bethe ansatz solution, it is possible to derive form factors and correlation functions. Details are given in 
\cite{jrlreview}.
~~\\

\centerline{{\bf Acknowledgements}}
~\\
AF and GS would like to thank S. R. Dahmen for discussions and CNPq-Conselho Nacional de Desenvolvimento
Cient\'{\i}fico e Tecnol\'ogico for financial support. AT thanks FAPERGS-Funda\c{c}\~ao de 
Amparo \`a Pesquisa do Estado do Rio Grande do Sul for financial support. 
JL gratefully acknowledges funding from the Australian Research Council and The University 
of Queensland through a Foundation Research Excellence Award.


\end{document}